\documentclass[aps,prl,twocolumn]{revtex4}

\usepackage{graphicx}

\usepackage{amssymb}
\usepackage{amsmath}
\usepackage{verbatim}
\usepackage{tabularx}

\usepackage{algorithmicx}
\usepackage{algpseudocode}

\begin{document}

\title{Renormalization-Group Geometry of Homeostatically Regulated Reentry Networks}

\author{Byung Gyu Chae}

\affiliation{Electronics and Telecommunications Research Institute, 218 Gajeong-ro, Yuseong-gu, Daejeon 34129, Republic of Korea}

\begin{abstract}

Reentrant computation--recursive self-coupling in which a network
continuously reinjects and reinterprets its own internal state--plays a
central role in biological cognition but remains poorly characterized
in neural network architectures.
We introduce a minimal continuous-time formulation of a homeostatically
regulated reentrant network (FHRN) and show that its population dynamics
admit an exact reduction to a one-dimensional radial flow.
This reduction reveals a dynamically fixed threshold for sustained
reflective activity and enables a complete renormalization-group (RG)
analysis of the reentry-homeostasis interaction.
We derive a closed RG system for the parameters governing structural
gain, homeostatic stiffness, and reentrant amplification, and show that
all trajectories are attracted to a critical surface defined by
$\gamma\rho=1$, where intrinsic leak and reentrant drive exactly balance.
The resulting phase structure comprises quenched, reactive, and
reflective regimes and exhibits a mean-field critical onset with universal scaling.
Our results provide an RG-theoretic characterization of reflective
computation and demonstrate how homeostatic fields stabilize deep
reentrant transformations through scale-dependent self-regulation.
\end{abstract}

\maketitle

\emph{Introduction}---Reentry--recursive bidirectional signaling between 
neural populations--is widely regarded as a core mechanism underlying
perception, memory, and reflective thought~\cite{1,2}.  
Classical computational models capture important aspects of recurrence
but do not implement reentry in the strict sense of Edelman's or Tononi's
theories.
Hopfield networks provide associative memory but lack a mechanism for
routing the current state back into the computation with
state-dependent gain~\cite{3,4}.  
Oja-type normalization rules regulate activity magnitudes without
supporting recursive reinterpretation of internal
representations~\cite{5,6}.  
Fast-weight architectures enable rapid context-dependent modulation
of synaptic efficacy~\cite{7,8,9,10,11,12}, and continuous-time recurrent
ODEs produce rich nonlinear dynamics~\cite{13,14,15}, yet these frameworks
act primarily as feed-forward transformations unrolled through time
rather than systems that explicitly \emph{re-enter} and reinterpret
their own latent states.

This conceptual gap motivates the formulation of a model in which
reentrant feedback is introduced as a \emph{distinct dynamical
operation} rather than an implicit byproduct of recurrence.
In biological terms, this corresponds to a cortical loop that returns a
population activity through a structured operation and modulates it via
population-level homeostasis~\cite{16,17,18,19,20}.  
In mathematical terms, it yields a dynamical architecture in which the
strength, stability, and scale-dependence of reentry can be analyzed
explicitly, allowing us to ask under what conditions reentry amplifies
or suppresses activity and how homeostatic mechanisms regulate this
balance across scales.

A recent continuous-time formulation of the Fast-Weights Homeostatic
Reentry Network (FHRN) addressed part of this question by deriving an
intrinsic neural ODE with explicit leak, spectral reentry, and radial
homeostatic damping~\cite{21,22}.  
That work established a quartic Lyapunov potential, a stable ring
attractor, and explicit Jacobian conditions, providing a dynamical
interpretation of regulated reentry distinct from classical RNNs,
fast-weight models, and liquid networks.
However, the analysis remained fundamentally \emph{local}: it described
behavior near fixed points or along the invariant homeostatic manifold,
but did not address how reentrant feedback behaves under changes of
scale, nor whether reflective computation admits a form of universality
analogous to critical phenomena.

In this work, we extend the FHRN framework from local dynamics to global
scale dependence by developing a renormalization-group (RG)
formulation of reentrant neural computation~\cite{23,24}.  
Starting from the continuous-time intrinsic equation
\begin{equation}
    \dot y = -y + \gamma W\, g(\|y\|)\, y ,
\end{equation}
we show that the representation-space radius $r=\|y\|$ provides an exact
reduction and allows the RG scale to be defined as $\ell=\ln r$.
Under this coarse-graining, the bare reentry gain $\gamma$, homeostatic
stiffness $\kappa$, and spectral radius $\rho$ acquire nontrivial scale
dependence.
Remarkably, their flows close into a three-dimensional RG system whose
geometry reveals a two-dimensional \emph{critical surface}
\begin{equation}
    \gamma \rho = 1,
    \label{eq:crit_new}
\end{equation}
on which reentrant amplification and homeostatic suppression exactly
balance.

We further show that this surface is infrared-attractive: large-scale
representations naturally evolve toward it, implying that reflective
reentry is not fine-tuned but emerges as a universal large-scale
property of homeostatically regulated recurrent computation.
This RG structure parallels dynamical renormalization phenomena in
nonlinear oscillators, the logistic map, and Kuramoto-type
synchronization~\cite{25,26,27,28},
indicating that the FHRN constitutes a minimal neural system exhibiting
a genuine critical geometry.
Taken together, these results bridge associative-memory theory,
continuous-time neural dynamics, and renormalization-group concepts,
revealing that reflective computation arises as a
\emph{scale-invariant phenomenon} governed by a critical surface in
$(\gamma,\kappa,\rho)$ space.

\begin{figure}[ht!]
\includegraphics[scale=0.5, trim= 1.0cm 12.5cm 0cm 0cm]{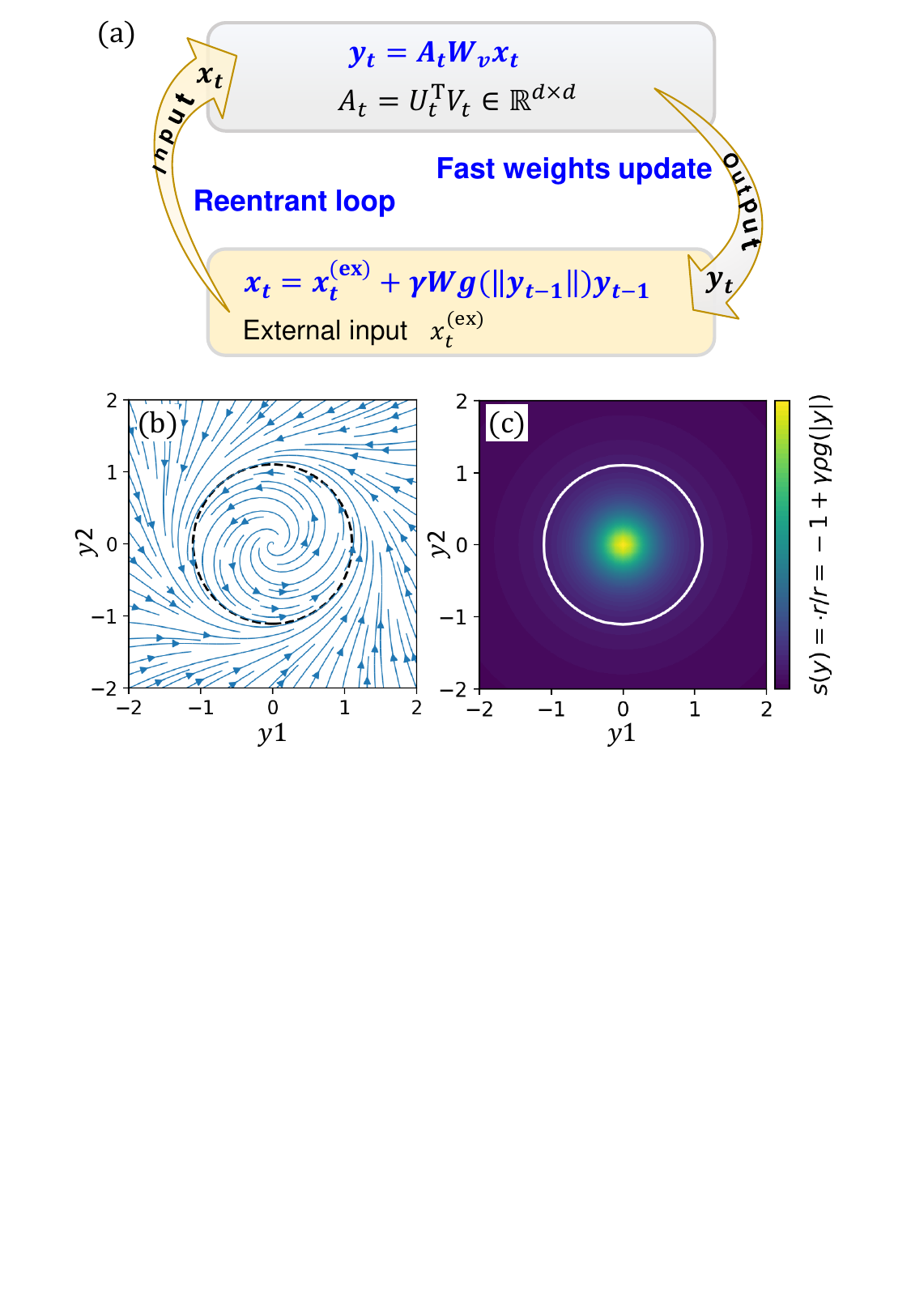}
\caption{
Homeostatically regulated reentrant dynamics and reflective shell formation.
(a) Schematic of the homeostatically regulated reentrant network.
External input $x_t^{(\mathrm{ex})}$ is combined with a reentrant signal
$\gamma W\, g(\|y_{t-1}\|)\, y_{t-1}$, producing the effective block input
$x_t$.
The population activity $y_t$ is computed through a fast-weight operator $A_t$,
while the reentrant loop feeds the internal state back into the next step with
homeostatic gain control.
(b) Phase portrait of the continuous-time FHRN dynamics in the reflective regime $\gamma\rho>1$.
Trajectories spiral under the reentry operator and converge toward a finite-radius
shell, indicating sustained reflective activity.
(c) Radial growth-rate map
$s(y)=\dot r/r=-1+\gamma\rho\, g(\|y\|)$, where $r=\|y\|$.
The white contour $s(y)=0$ identifies the homeostatic reflective shell.
Inside the shell ($s>0$) trajectories are radially unstable and expand outward,
while outside ($s<0$) they contract inward, rendering the shell globally attracting.
}
\end{figure}

\vspace{6pt}
\emph{Radial ODE for homeostatic reentry networks}—Our analysis
begins from the intrinsic neural dynamics
\[
    \dot y(t)
    = -y(t) + \gamma\, W\, g(\|y(t)\|)\, y(t),
\]
in which the scalar gain $\gamma$, the reentry operator $W$, and the
homeostatic field $g(\cdot)$ jointly determine whether population
activity is amplified or suppressed.
The reentry operator represents a structured reentry map learned within a
larger transformer-like architecture, while the nonlinear gain
$g(\cdot)$ implements population-level homeostatic suppression as the
activity norm grows.

We adopt the biologically motivated radial gain
\begin{equation}
    g(r)
    = \frac{1}{1+\kappa(r^2-1)}, \qquad \kappa>0,
\end{equation}
which penalizes deviations from unit activity and yields an exactly
solvable radial dynamics.
Figure~1 illustrates the resulting phase portraits of the full vector
dynamics for representative parameters.
Despite the absence of any explicit radial constraint, trajectories from
a broad set of initial conditions converge toward a finite-radius shell
when reentrant amplification dominates leakage.

Let $W v=\rho v$ denote the dominant eigenmode of the reentry operator.
Projecting the full system onto $v$ and writing $y=r v$ yields the exact
radial ODE
\begin{equation}
    \dot r
    = \left[-1+\gamma\rho\,g(r)\right] r .
    \label{eq:radial_new}
\end{equation}
This reduction captures the long-time behavior observed in Fig.~1:
angular degrees of freedom rapidly mix under $W$, while the slow radial
mode governs convergence toward or away from the homeostatic shell.
Equation~\eqref{eq:radial_new} therefore encapsulates the stability,
homeostatic regulation, and reentrant amplification properties of the
full system.

Stationary radii satisfy $\dot r=0$, yielding either $r=0$ or
\begin{equation}
    r_\infty^2
    = 1 + \frac{\gamma\rho-1}{\kappa}.
\end{equation}
A nontrivial stationary shell exists only when
\[
    \gamma\rho \ge 1 ,
\]
with the equality $\gamma\rho=1$ marking the reflective threshold.
Below the threshold the only stable attractor is $r=0$, so generic
trajectories decay to silence, whereas above it the dynamics converges
to a finite homeostatic shell whose radius is set by $\kappa$.
Numerically, we find that this shell remains globally attracting over a
wide range of homeostatic curvatures, demonstrating that
$\kappa$ modulates the shell radius without altering the existence of
the reflective phase (see the details in Appendix~A).

Although Eq.~\eqref{eq:crit_new} arises from a balance of deterministic
flows in the radial ODE, we show below that the same condition reappears
as an infrared-attractive critical surface under coarse-graining of the
radial dynamics.

\vspace{6pt}
\emph{Renormalization-group geometry}—To characterize the intrinsic scale dependence of the reentry–homeostasis
coupling, we apply a Wilsonian log-shell coarse-graining directly to the
exact radial dynamics
\begin{equation}
    \dot r = F(r;\lambda,\kappa)\, r,
    \qquad
    F(r;\lambda,\kappa)
    = -1 + \frac{\lambda}{1+\kappa(r^{2}-1)},
\end{equation}
where $\lambda=\gamma\rho$ denotes the composite reentry amplification.
Introducing the logarithmic radius $\ell=\ln r$, a shell transformation
$r=r'e^{\delta\ell}$ is applied to the \emph{full} radial equation.
Under this transformation the equation of motion becomes
$\dot r' = F(r'e^{\delta\ell};\lambda,\kappa)\,r'$, since the common
factor $e^{\delta\ell}$ cancels identically between the left- and
right-hand sides.
It is therefore sufficient to define an effective coarse-grained flow
$F_{\mathrm{eff}}(r')=F(r'e^{\delta\ell};\lambda,\kappa)$.
Requiring that $F_{\mathrm{eff}}$ retain the same functional form as
$F(r';\lambda',\kappa')$ then uniquely determines the Wilsonian flow of
the couplings under coarse-graining.

Expanding the exact flow about the homeostatic shell $r=1$ and writing
$s=r^{2}-1$, we obtain the local series
\begin{equation}
    F(r)
    = (\lambda-1)
      - \lambda\kappa\, s
      + \lambda\kappa^{2}s^{2}
      + O(s^{3}),
\end{equation}
where the constant term controls the growth rate on the shell and the
higher-order terms encode the curvature of the homeostatic suppression.
Under the shell transformation,
$s' = r'^{2}-1$ maps to
$s=(1+s')e^{2\delta\ell}-1$, and inserting this relation into the local
expansion yields the coarse-grained flow $F_{\mathrm{eff}}(r')$.
Matching the constant and linear coefficients of this flow to those of
$F(r';\lambda',\kappa')$ fixes the \emph{raw} Wilsonian RG increments
$\lambda'=\lambda+\beta_\lambda\,\delta\ell$ and
$\kappa'=\kappa+\beta_\kappa\,\delta\ell$.

Carrying out this matching to leading order in $\delta\ell$, while
retaining the curvature contribution proportional to $s^{2}$, yields
the raw beta functions
\begin{equation}
    \beta_\lambda = -2\lambda\kappa,
    \qquad
    \beta_\kappa = 2\kappa(1-\kappa),
\end{equation}
which follow uniquely from the Wilsonian log-shell construction and
involve no additional ansatz.
Writing $\Delta=\lambda-1$, the corresponding flow of the deviation from
the reflective threshold is
\begin{equation}
    \beta_\Delta^{\mathrm{raw}}
    = -2\kappa(1+\Delta).
\end{equation}

To analyze the near-critical structure, we reorganize the raw flows in a
normal-form (interaction) RG scheme adapted to the critical manifold
$\Delta=0$.
Removing the canonical contribution that survives at $\Delta=0$ defines
the interaction beta function
\begin{equation}
    \beta_\Delta^{\mathrm{int}} = -2\kappa\,\Delta + O(\Delta^{2}),
\end{equation}
which vanishes identically on the reflective manifold.
In this interaction scheme, $\Delta$ is an irrelevant direction: for
$\kappa>0$, deviations from the reflective threshold decay
exponentially under coarse-graining.
Consequently, the surface
\begin{equation}
    \mathcal S_c
    = \{(\gamma,\kappa,\rho)\mid \gamma\rho=1\}
\end{equation}
is an invariant RG manifold.

In contrast, the raw flow of the homeostatic stiffness $\kappa$ is purely
canonical.
Allowing for interaction corrections consistent with analyticity,
invariance of $\mathcal S_c$, and boundedness of $\kappa\in[0,1]$, the
leading-order interaction contribution takes the form
\begin{equation}
    \beta_\kappa
    = 2\kappa(1-\kappa)
      - e\,\Delta\,\kappa(1-\kappa)
      + O(\Delta^{2}),
\end{equation}
where $e>0$ is an architecture-dependent coefficient.

Because $\lambda=\gamma\rho$, the RG flow of the composite amplification
obeys the exact identity
\[
    \beta_\lambda
    = \rho\,\beta_\gamma + \gamma\,\beta_\rho .
\]
This relation fixes $\beta_\lambda$ uniquely through the Wilsonian
log-shell calculation, but leaves a scheme freedom in its decomposition
into the primitive couplings $\beta_\gamma$ and $\beta_\rho$.
Here we adopt an interaction RG scheme in which the canonical rescaling
contribution to $\beta_\kappa$ is absorbed, so that only the
interaction-induced flow is retained.

A convenient one-parameter family of RG decompositions that reproduces
the Wilsonian flow of $\lambda$ exactly is
\begin{align}
    \beta_\gamma
    &= -a\,\kappa\,\frac{\gamma\rho-1}{\rho},
    \\
    \beta_\rho
    &= -(2-a)\,\kappa\,\frac{\gamma\rho-1}{\gamma},
    \\
    \beta_\kappa
    &= - e\,(\gamma\rho-1)\,\kappa(1-\kappa),
\end{align}
where $a\in\mathbb{R}$ parameterizes the choice of RG coordinates in the
$(\gamma,\rho)$ plane, and $e>0$ encodes the leading interaction-induced
modulation of the homeostatic stiffness.

All members of this family generate the same Wilsonian flow for the
physically relevant composite coupling $\lambda$, differing only by a
scheme-dependent redistribution between $\gamma$ and $\rho$.
In this interaction scheme, the flow of $\kappa$ vanishes identically on
the reflective manifold $\gamma\rho=1$, indicating that homeostatic
stiffness becomes scale-invariant precisely at criticality.

Taken together, the coupled flows of
$(\gamma,\kappa,\rho)$ define a three-dimensional RG system driven by the
reflective deviation $\Delta=\gamma\rho-1$.
For $\kappa>0$, trajectories are attracted toward the critical manifold
$\mathcal S_c$, demonstrating that the microscopic reflective threshold
emerging from the deterministic radial dynamics is promoted by
coarse-graining into a universal RG structure.
Full details of the Wilsonian matching and normal-form construction are
given in Appendix~B.

On the other hand, the three computational regimes follow directly from
the existence and stability properties of the fixed points of the radial ODE in Eq~\eqref{eq:radial_new}.
Linearizing about the origin yields
$\dot r \approx [-1+\gamma\rho/(1-\kappa)]r$, showing that $r=0$ is
stable when $\gamma\rho < 1-\kappa$, which defines the quenched--decay
boundary where leakage dominates amplification.
A nonzero stationary radius satisfies
$-1+\gamma\rho\,g(r_\infty)=0$, which requires $\gamma\rho>1$ and yields
a globally attracting homeostatic shell, identifying the reflective
threshold.

Between these limits, in the intermediate region
$1-\kappa < \gamma\rho < 1$, the dynamics admits transient or
metastable amplification and may support finite-radius activity for
extended times depending on initial conditions.
However, the nonzero fixed point is not globally stable, and generic
trajectories ultimately relax toward silence.
This region therefore defines the reactive regime.

Together, these inequalities partition the $(\gamma,\kappa,\rho)$ space
into three computational phases: a \textit{quenched} phase
($\gamma\rho < 1-\kappa$) in which activity rapidly decays to zero; a
\textit{reactive} regime ($1-\kappa < \gamma\rho < 1$) characterized by
transient or metastable amplification without a globally attracting
shell; and a \textit{reflective} phase ($\gamma\rho > 1$) in which
trajectories converge to a stable homeostatic shell and sustain
reentrant activity.

Fixed-$\rho$ slices reveal boundaries as the simple curves
$\gamma\rho = 1$ and $\gamma\rho = 1-\kappa$, while fixed-$\kappa$
sections display hyperbolic geometry.
The global RG flow forms an attractive funnel toward the
critical surface $\mathcal S_c$, across which reflective computation
becomes a scale-invariant property of the network.

\begin{figure}[ht!]
\includegraphics[scale=0.55, trim= 0.2cm 15.5cm 0cm 0cm]{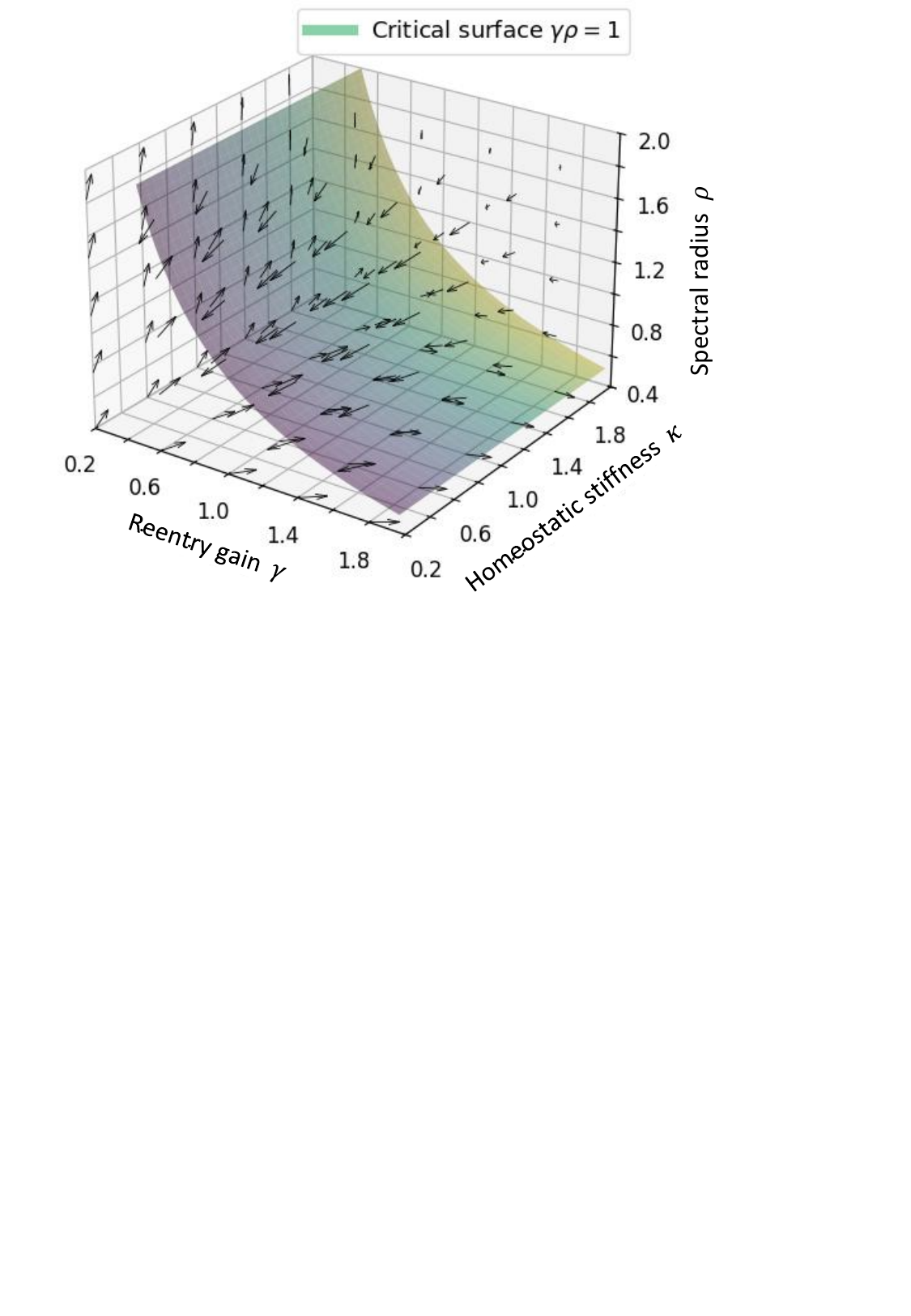}
\caption{
Renormalization-group flow of the homeostatically regulated reentrant
network in the full $(\gamma,\kappa,\rho)$ parameter space.
Arrows indicate the normalized RG vector field
$(\beta_\gamma,\beta_\kappa,\beta_\rho)$, emphasizing the topology of the
flow independently of its local speed.
The translucent surface marks the critical manifold $\gamma\rho=1$,
which is approached transversely by the RG trajectories and is therefore
infrared-attractive.
This surface separates quenched and reflective regimes and unifies
reentrant amplification, homeostatic regulation, and spectral scaling
into a single scale-invariant geometric structure.
}
\end{figure}

The full three-dimensional RG flow generated by Eqs. (13)-(15) is visualized in Fig.~2.
The critical surface $\gamma\rho = 1$ appears as a smooth, curved sheet
in $(\gamma,\kappa,\rho)$ space, separating reflective and
non-reflective regimes.
Vector arrows indicate the local RG velocity field: for all trajectories
with $\kappa>0$, the flow is directed toward the surface, demonstrating
that it is infrared-attractive.

For small $\kappa$, the dynamics are dominated by reentry amplification
through $\gamma$ and $\rho$, while increasing $\kappa$ enhances the
damping of deviations from the reflective condition.
The resulting funnel-like geometry shows that coarse-graining drives the
effective coupling $\gamma\rho$ toward unity over a wide range of
initial conditions.
This global structure confirms that reflective computation corresponds
to a scale-invariant critical manifold rather than a fine-tuned line in
parameter space.

Figure~3(a) shows a two-dimensional slice of the RG flow in the
$(\gamma,\kappa)$ plane at fixed spectral radius $\rho=1.0$.
Two boundaries organize the dynamics: the dashed line
$\gamma\rho = 1-\kappa$, separating quenched and reactive regimes, and
the vertical line $\gamma\rho = 1$ marking the reflective threshold.
The RG vector field indicates that trajectories in the reactive region
are driven predominantly toward increasing reentry gain $\gamma$ and
converge toward the critical boundary $\gamma\rho = 1$, while trajectories
in the quenched region flow toward enhanced homeostatic regulation.
Beyond the reflective threshold $\gamma\rho>1$, the flow stabilizes on
the reflective side of the phase diagram.
This slice shows that, for fixed intrinsic structure $\rho$, coarse-graining
systematically adjusts reentrant amplification to balance homeostatic
suppression.

Figure~3(b) presents the complementary RG slice in the $(\gamma,\rho)$
plane at fixed homeostatic stiffness $\kappa=1$.
The critical curve $\gamma\rho=1$ appears as a hyperbola separating
subcritical and reflective regimes.
The RG vector field shows that trajectories from both sides are attracted
toward this curve, confirming that it constitutes an infrared-stable
manifold.
For $\gamma\rho<1$, coarse-graining suppresses amplification by relaxing
either $\gamma$ or $\rho$, while for $\gamma\rho>1$ the flow reduces the
effective gain until the two balance.
Thus, although $\gamma$ and $\rho$ represent distinct mechanisms—bare
reentry feedback and structural amplification—they renormalize
cooperatively to preserve the scale-invariant reflective threshold.

\begin{figure}[ht!]
\includegraphics[scale=0.5, trim= 0.7cm 20cm 0cm 0cm]{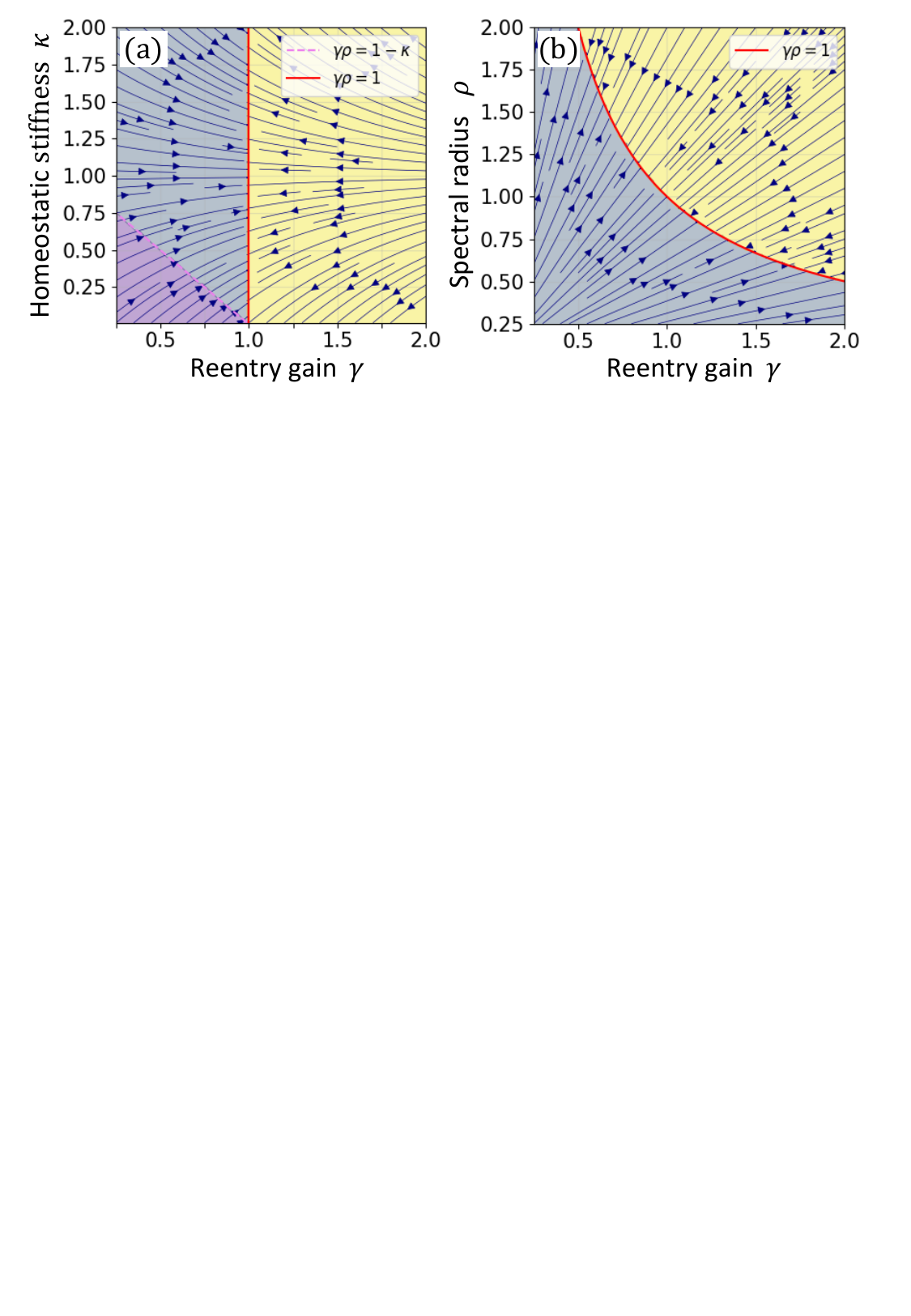}
\caption{
Two-dimensional slices of the RG flow highlighting the phase structure.
Arrows denote the normalized RG vector field, emphasizing flow topology
rather than local speed.
Background colors indicate quenched (blue), reactive (purple), and
reflective (yellow) regimes.
(a) RG streamlines in the $(\gamma,\kappa)$ plane at fixed $\rho=1$.
The vertical red line marks the reflective critical condition
$\gamma\rho=1$, while the dashed line $\gamma\rho=1-\kappa$ separates the
quenched and reactive regions.
Flows approach the critical line transversely, demonstrating its
infrared-attractive character.
(b) RG streamlines in the $(\gamma,\rho)$ plane at fixed $\kappa$.
The red curve $\gamma\rho=1$ defines the reflective manifold, which again
acts as an infrared attractor separating quenched and reflective phases.
}
\end{figure}

\vspace{6pt}
\emph{Stability of the critical surface}—To assess the stability of the
critical surface, we linearize the RG equations around an arbitrary
point on $\mathcal{S}_c$.
Because each $\beta$-function is proportional to the composite deviation
$\Delta=\gamma\rho-1$, the Jacobian evaluated on the surface takes the
outer-product form
\[
J_{ij}
= \left. \frac{\partial\beta_i}{\partial x_j}\right|_{\Delta=0}
= C_i\,(\rho\,\delta_{\gamma j} + \gamma\,\delta_{\rho j}),
\qquad
x_j\in\{\gamma,\kappa,\rho\},
\]
with coefficients
$C_\gamma=-a\kappa/\rho$,
$C_\kappa=-e\kappa(1-\kappa)$,
and $C_\rho=-(2-a)\kappa/\gamma$,
evaluated on $\mathcal{S}_c$.
This structure implies two vanishing eigenvalues corresponding to
perturbations tangent to the critical surface (See the details in Appendix C).
The remaining direction is normal to $\mathcal{S}_c$ and is governed by
the flow of the composite deviation
$\Delta=\gamma\rho-1$.
Using the chain rule,
\begin{equation}
    \dot\Delta
    = \rho\,\dot\gamma + \gamma\,\dot\rho
    = -\kappa(a+2-a)\,\Delta
    = -2\kappa\,\Delta ,
\end{equation}
which yields a single transverse eigenvalue
\begin{equation}
    \lambda_{\perp} = -2\kappa < 0 .
\end{equation}
Hence $\mathcal{S}_c$ is strictly infrared-attractive: perturbations that
modify the composite amplification $\gamma\rho$ decay exponentially
under coarse-graining, while directions tangent to the surface remain
marginal.

This stability structure matches the global geometry shown in Fig.~2.
RG trajectories funnel toward $\mathcal{S}_c$ along the unique stable
direction associated with $\Delta$, confirming that reflective balance is
a scale-invariant property of the FHRN dynamics.

\vspace{6pt}
\emph{Critical scaling and order-parameter exponent---}The reflective phase of the FHRN admits a natural order parameter measuring
the departure of the steady-state activity from the homeostatic shell.  
The stationary radius in the reflective regime satisfies $ r_\infty^2$ in Eq. (5),
suggesting the definition
$m := r_\infty - 1$.
Approaching the dynamical critical line $\gamma_c = 1/\rho$ from above,
the fixed point expands as
\begin{equation}
    r_\infty
    = 1 + \frac{\gamma\rho - 1}{2\kappa}
      + \mathcal{O}\!\left((\gamma\rho - 1)^2\right),
\end{equation}
yielding the mean-field scaling law
\begin{equation}
    m \;\propto\; (\gamma - \gamma_c)^{\beta_{\mathrm{op}}},
    \qquad \beta_{\mathrm{op}} = 1.
\end{equation}

The exponent $\beta_{\mathrm{op}} = 1$ reflects the linear
bifurcation of the reflective radius away from the homeostatic shell as the
reentry gain crosses threshold, matching the classical mean-field onset in
saddle-node and $\phi^4$-type theories.  
Numerical steady-state solutions confirm the analytic prediction:
$m$ exhibits a clean power-law dependence on $\gamma - \gamma_c$ with slope
$1$ on log--log axes in Fig. 4.

In addition, linearization of the radial dynamics shows that the relaxation
time diverges as $\tau_r \sim \kappa^{-1}$ (See the details in Appendix D), so that the dynamical approach to
the steady state shares the same mean-field scaling structure associated with
the onset of reflective activity.  
Together with the RG flow in the $(\gamma,\kappa)$ plane, these results
establish the reflective phase as a genuine critical regime with universal
onset exponent $\beta_{\mathrm{op}} = 1$.

\begin{figure}[ht!]
\includegraphics[scale=0.55, trim= 0.2cm 19cm 0cm 0cm]{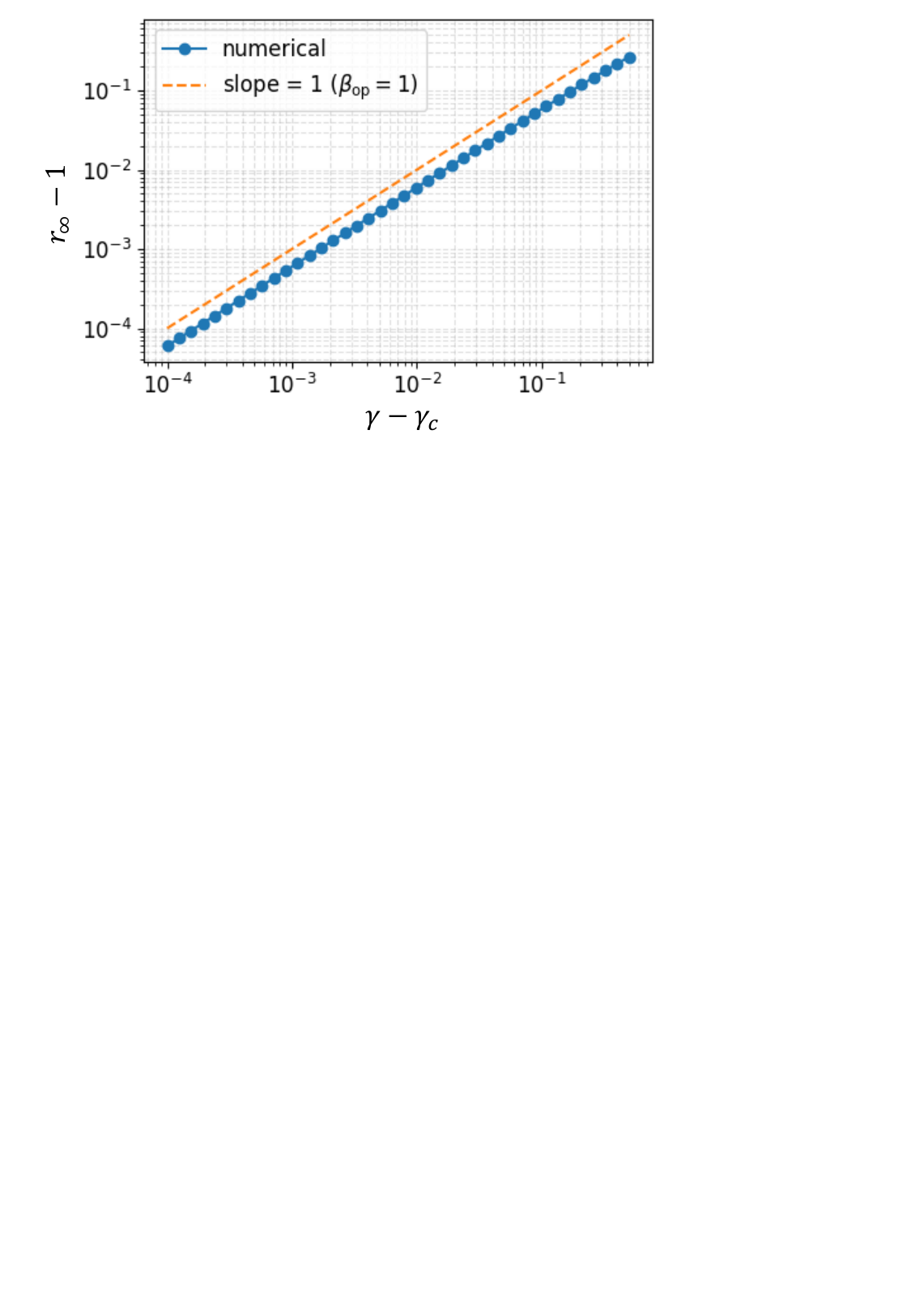}
\caption{
Critical scaling of the order parameter $m=r_\infty-1$ near $\gamma_c=1/\rho$.
Numerical solutions follow the predicted mean-field exponent $\beta_{\mathrm{op}}=1$ over several decades.
}
\end{figure}

\vspace{6pt}
\emph{Discussion}---We have developed a dynamical and RG-theoretic
framework for homeostatically regulated reentrant networks.
The FHRN admits an exact reduction to a one-dimensional radial flow,
revealing a dynamically fixed threshold for reflective computation.
Wilsonian coarse-graining of this flow shows that the threshold extends
to a two-dimensional infrared-attractive manifold $\gamma\rho=1$ in
$(\gamma,\kappa,\rho)$ space, unifying reentrant amplification,
homeostatic regulation, and spectral structure within a single
scale-dependent geometry.

This structure demonstrates that reflective computation is not a
fine-tuned phenomenon but a scale-invariant attractor of the dynamics.
At the level of neural activity, contraction toward a stable
homeostatic shell guarantees bounded reentrant amplification.
At the level of renormalization, the composite gain $\gamma\rho$ flows
toward the critical surface, indicating that deviations from reflective
balance constitute an irrelevant RG direction.
The agreement between these two notions of stability—convergence
$r\!\to\!r_\infty$ in state space and $\gamma\rho\!\to\!1$ in parameter
space—shows that reflective computation behaves as a genuine dynamical
fixed structure.
Numerical phase portraits further confirm that this shell remains
globally attracting across a wide range of homeostatic curvatures
$\kappa$, indicating that the existence of the reflective phase is
structurally robust rather than a consequence of precise parameter
tuning.

Below the reflective threshold, the microscopic dynamics may still
admit nonzero attracting radii for particular choices of the
homeostatic nonlinearity.
Such inner attractors arise from nonuniversal features of the gain
function and do not persist under coarse-graining.
Accordingly, they should be interpreted as reactive or transient
structures rather than reflective phases in the RG sense.

Within this RG framework the three control parameters play distinct
scaling roles.
The composite reentry gain $\gamma\rho$ defines the single relevant
direction that controls departure from the reflective manifold, while
the homeostatic strength $\kappa$ governs only the local curvature of
the radial flow and therefore spans an irrelevant (or weakly marginal)
direction.
Consistent with this interpretation, varying $\kappa$ shifts the radius
of the attracting shell without altering the critical surface or
introducing additional fixed structures, confirming that $\kappa$
controls nonuniversal amplitudes rather than the universality class
itself.
The internal feedback scale $\rho$ enters solely through a uniform
rescaling of the reentrant signal, generating a marginal direction
associated with spectral normalization.
This one-relevant, one-irrelevant, and one-marginal structure fully
characterizes the RG geometry underlying the FHRN.

The resulting RG flow places the FHRN within a mean-field universality class.
The onset of reflectivity exhibits an order-parameter exponent
$\beta_{\mathrm{op}}=1$, and the deviation flow is structurally
analogous to the dynamical RG governing logistic-map bifurcations and
Kuramoto synchronization thresholds \cite{26,28}.
Importantly, the robustness of the reflective shell under variation of
homeostatic curvature indicates that this mean-field scaling persists
beyond idealized isotropic limits, i.e., it remains unchanged under variations of spectral anisotropy 
and homeostatic curvature that deform local dynamics without altering the critical manifold..
These analogies suggest that reflective processing in biological and
artificial recurrent networks may follow universal scaling laws arising
from the balance between leakage and structured reentrant drive.

Finally, the FHRN framework can be incorporated into transformer-like
architectures by treating the reentry operator as a learnable feedback
map acting on internal representations.
From an RG perspective, gradient-based training may naturally steer
parameters toward the reflective surface $\gamma\rho=1$, stabilizing
deep reentrant computation while preserving sensitivity to external
inputs.

\vspace{6pt}
\emph{Conclusion}---We presented a unified dynamical and RG-theoretic description of
homeostatically regulated reentrant computation.
The exact reduction to a radial flow identifies $\gamma\rho=1$ as the
onset of sustained reflectivity, while RG coarse-graining shows that this
threshold forms an infrared-attractive critical manifold.
These results demonstrate that reflective computation emerges as a
scale-invariant attractor characterized by a mean-field onset exponent
$\beta_{\mathrm{op}} = 1$ and a single relevant deviation mode.
Because the reentry operator can be trained inside modern
transformer architectures, our framework suggests that learned models may
naturally evolve toward the reflective manifold during optimization,
stabilizing the reentrant transformations while enhancing deep reflective processing.

\vspace{6pt}
\emph{Acknowledgements}---This work was partially supported by the Institute of Information \& Communications Technology Planning \& Evaluation (IITP) grant 
funded by the Korea government (MSIT) (IITP-RS-2025-02214780).

The author acknowledges the support of ChatGPT (GPT-5, OpenAI) for assistance in literature review and conceptual structuring during early development.

\clearpage
\appendix

\renewcommand{\thefigure}{S\arabic{figure}}
\renewcommand{\theequation}{S\arabic{equation}}

\setcounter{figure}{0}
\setcounter{equation}{0}

\vspace*{1.5cm}
{\centering\large\bfseries End Matter\par}
\vspace{1.0cm}

\section{Appendix A: Robustness with respect to homeostatic curvature}
\label{app:kappa}

\begin{figure}[ht!]
\includegraphics[scale=0.5, trim= 0.7cm 12.5cm 0cm 0cm]{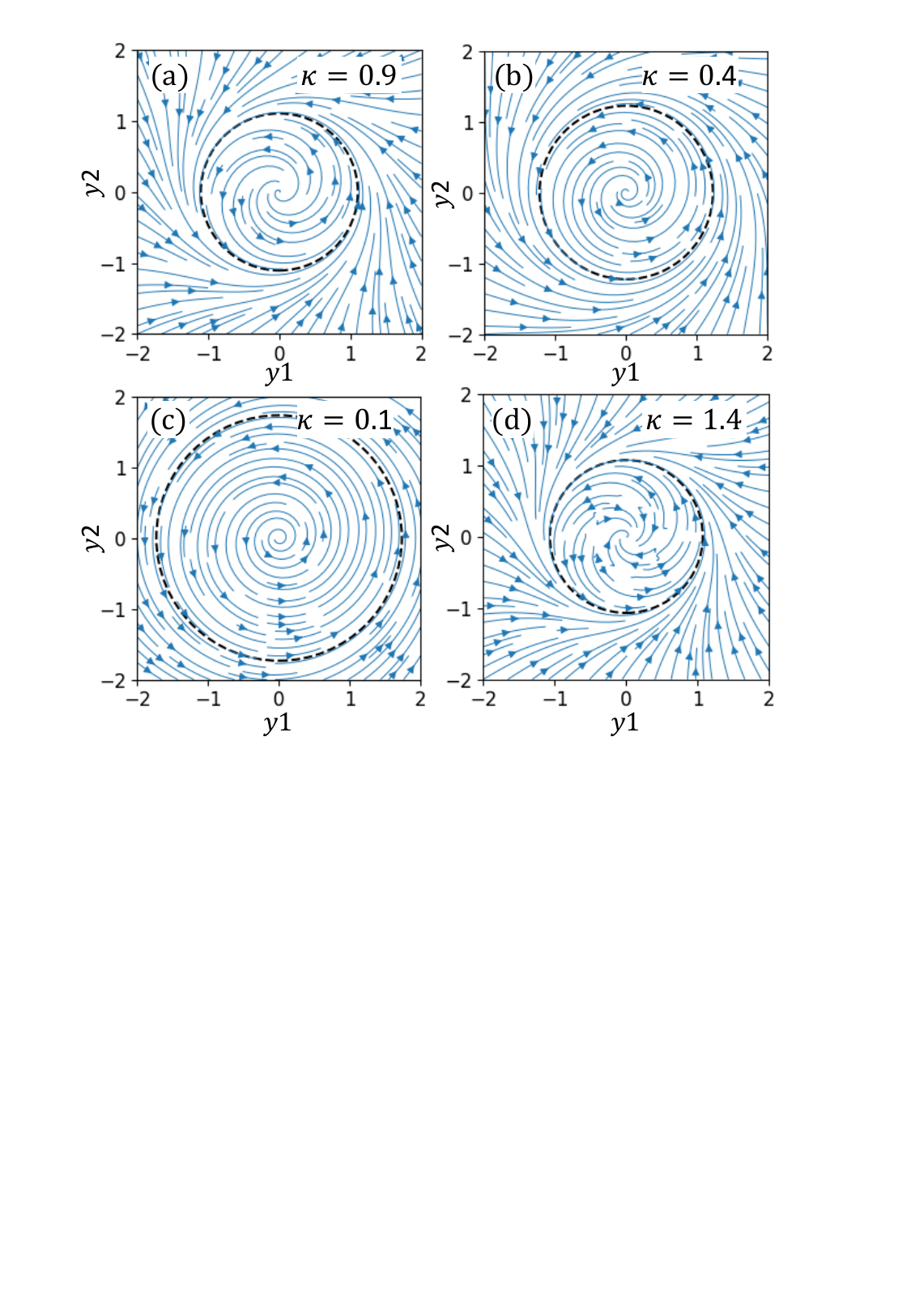}
\caption{
Dependence of the homeostatic reflective shell on the curvature parameter
$\kappa$.
Phase portraits of the full FHRN dynamics
are shown for fixed reentry gain $\gamma=1.2$ and spectral radius $\rho=1$,
while varying the homeostatic curvature $\kappa$.
Dashed circles indicate the stationary shell radius
$r_\infty^2 = 1 + (\gamma\rho-1)/\kappa$ predicted by the radial theory.
Panels (a)--(c) show that decreasing $\kappa$ expands the reflective shell
while preserving global convergence and spiral dynamics.
Panel (d) illustrates that for larger $\kappa$ the shell contracts but
remains stable.
}
\end{figure}

To assess the robustness of the reflective phase beyond the representative
parameter choices used in the main text, we systematically varied the
homeostatic curvature $\kappa$ while keeping the reentry gain $\gamma$
and dominant reentry eigenvalue $\rho$ fixed in the reflective regime
$\gamma\rho>1$.

Figure S1 displays representative phase portraits of
the full vector dynamics
\[
    \dot y = -y + \gamma\, W\, g(\|y\|)\, y
\]
for several values of $\kappa$.
Across a broad range of homeostatic strengths, trajectories from diverse
initial conditions converge to a single attracting shell, indicating
that the existence of the reflective phase does not depend on fine-tuned
choices of $\kappa$.

As $\kappa$ is reduced, the radius of the attracting shell increases in
quantitative agreement with the stationary solution of the effective
radial equation,
\[
    r_\infty^2 = 1 + \frac{\gamma\rho - 1}{\kappa}.
\]
Conversely, increasing $\kappa$ leads to a contraction of the shell,
while leaving its stability and basin of attraction intact.
No additional attractors, bifurcations, or basin fragmentation are
observed over the range of $\kappa$ explored, demonstrating global
convergence to a single reflective shell.

From a renormalization-group perspective, these observations support the
interpretation of $\kappa$ as an irrelevant (or weakly marginal)
parameter.
While $\kappa$ modifies the local curvature of the radial flow and sets
the nonuniversal amplitude of the reflective state, it does not shift the
critical surface $\gamma\rho=1$ nor alter the order-parameter scaling.
The persistence of the reflective shell under wide variation of
$\kappa$ therefore provides direct numerical evidence for the structural
stability of the FHRN universality class.

\section{Appendix B: Wilsonian log-shell RG for $\gamma$ and $\kappa$ in the FHRN model}

We consider the exact radial dynamics of the FHRN,
\begin{equation}
  \dot r = F(r;\lambda,\kappa)\,r,
  \qquad
  F(r;\lambda,\kappa)
  = -1 + \frac{\lambda}{1+\kappa(r^2-1)},
\end{equation}
where the composite amplification is defined as
\begin{equation}
  \lambda := \gamma\rho,
  \qquad
  \Delta := \lambda - 1,
  \qquad
  s := r^2 - 1 .
\end{equation}

A Wilsonian RG step is implemented as a logarithmic shell rescaling of
the radial coordinate,
\begin{equation}
  r = r' e^{\delta\ell},
  \qquad
  s' := r'^2 - 1 ,
\end{equation}
with $\delta\ell>0$ infinitesimal.
Importantly, this transformation is applied to the \emph{full equation
of motion}.
Under the rescaling one finds
\[
  \dot r
  = \frac{d}{dt}\!\left(e^{\delta\ell} r'\right)
  = e^{\delta\ell}\dot r',
  \qquad
  F(r;\lambda,\kappa)\,r
  = F(r'e^{\delta\ell};\lambda,\kappa)\,e^{\delta\ell} r' ,
\]
so that the common factor $e^{\delta\ell}$ cancels identically and the
coarse-grained dynamics takes the form
\begin{equation}
  \dot r' = F(r'e^{\delta\ell};\lambda,\kappa)\,r'.
\end{equation}
Thus defining an effective flow
\begin{equation}
  F_{\mathrm{eff}}(r')
  := F(r'e^{\delta\ell};\lambda,\kappa)
\end{equation}
is fully equivalent to applying the shell transformation to the entire
ODE.  The RG step therefore amounts to determining how the functional
form of $F$ is preserved under this rescaling by an induced flow of the
couplings $(\lambda,\kappa)$.

In what follows we expand systematically in $s'$ and $\delta\ell$ to the
order required to obtain the $\kappa^{2}$ correction to the
Wilsonian $\beta$-function of $\kappa$.

Using the geometric series
\begin{equation}
  \frac{1}{1+\kappa s}
  = 1 - \kappa s + \kappa^2 s^2 + O(s^3),
\end{equation}
the exact flow can be written locally as
\begin{align}
  F(s)
  &= -1 + \lambda\left(1 - \kappa s + \kappa^2 s^2 + O(s^3)\right) \\
  &= \Delta - \lambda\kappa\, s + \lambda\kappa^2\, s^2 + O(s^3).
\end{align}
This expansion is purely kinematic and follows directly from the
definition of $F$.

\vspace{10pt}
\emph{Log-shell transformation for $s$ and the $s^2$ cross-term\rm{:}}
From $r = r' e^{\delta\ell}$ we obtain
\begin{align}
  s
  &= r^2 - 1
   = r'^2 e^{2\delta\ell} - 1
   = (1+s')e^{2\delta\ell} - 1 .
\end{align}
Expanding $e^{2\delta\ell}$ to first order gives
\begin{equation}
  s = s' + 2\delta\ell + 2 s'\delta\ell + O(\delta\ell^2).
  \label{eq:s_transform_first}
\end{equation}

Squaring \eqref{eq:s_transform_first} yields
\begin{align}
  s^2
  &= \left(s' + 2\delta\ell + 2s'\delta\ell\right)^2 + O(\delta\ell^2) \\
  &= (s')^2 + 4s'\delta\ell + 4(\delta\ell)^2
     + O(\delta\ell s'^2,\delta\ell^2).
\end{align}
For the extraction of the coefficient linear in $s'$, the only relevant
term is
\begin{equation}
  s^2 \supset 4s'\delta\ell .
  \label{eq:s2_cross}
\end{equation}

\vspace{6pt}
\emph{Coarse-grained effective flow and coefficients $c_0',c_1'$\rm{:}}
The coarse-grained flow is defined as
\begin{equation}
  F_{\mathrm{eff}}(r')
  := F(r' e^{\delta\ell};\lambda,\kappa)
  = F\bigl(s(s',\delta\ell);\lambda,\kappa\bigr).
\end{equation}
Substituting the expansions above, we obtain
\begin{align}
\begin{split}
  F_{\mathrm{eff}}(r')
  &= \Delta - \lambda\kappa\, s + \lambda\kappa^2\, s^2 + O(s'^2,\delta\ell^2) \\
  &= \Delta
     - \lambda\kappa\left(s' + 2\delta\ell + 2s'\delta\ell\right)
     + \lambda\kappa^2\left(4s'\delta\ell\right) \\
     &+ O(s'^2,\delta\ell^2) \\
  &= \left(\Delta - 2\lambda\kappa\,\delta\ell\right)
   + \left[-\lambda\kappa(1+2\delta\ell)
           + 4\lambda\kappa^2\delta\ell\right] s' \\
   &+ O(s'^2,\delta\ell^2).
\end{split}
\end{align}
Therefore,
\begin{equation}
\begin{split}
  &c_0' = \Delta - 2\lambda\kappa\,\delta\ell + O(\delta\ell^2),
  \qquad \\
  &c_1' = -\lambda\kappa
         + \left(-2\lambda\kappa + 4\lambda\kappa^2\right)\delta\ell
         + O(\delta\ell^2).
\end{split}
\end{equation}

\vspace{6pt}
\emph{Wilsonian matching and raw beta functions\rm{:}}
We impose matching to the same functional form,
\begin{equation}
  F_{\mathrm{eff}}(r') \equiv \Delta' - \lambda'\kappa' s' + O(s'^2),
\end{equation}
with
\begin{equation}
  \lambda' = \lambda + \beta_\lambda\,\delta\ell,
  \qquad
  \kappa' = \kappa + \beta_\kappa\,\delta\ell,
  \qquad
  \Delta' = \Delta + \beta_\Delta\,\delta\ell ,
\end{equation}
where $\beta_\Delta \equiv \beta_\lambda$.

From $\Delta' = c_0'$ we obtain
\begin{equation}
  \beta_\Delta
  = \frac{d\Delta}{d\ell}
  = -2\lambda\kappa
  = -2\kappa(1+\Delta),
\end{equation}
or equivalently
\begin{equation}
  \beta_\lambda
  = \frac{d\lambda}{d\ell}
  = -2\lambda\kappa .
\end{equation}

From $-\lambda'\kappa' = c_1'$ we find
\begin{align}
  -\lambda'\kappa'
  &= -(\lambda+\beta_\lambda\delta\ell)(\kappa+\beta_\kappa\delta\ell)
     + O(\delta\ell^2) \\
  &= -\lambda\kappa
     -(\lambda\beta_\kappa+\kappa\beta_\lambda)\delta\ell
     + O(\delta\ell^2).
\end{align}
Matching the $\delta\ell$ coefficient gives
\begin{equation}
  \lambda\beta_\kappa + \kappa\beta_\lambda
  = 2\lambda\kappa - 4\lambda\kappa^2 .
\end{equation}
Using $\beta_\lambda=-2\lambda\kappa$, we obtain
\begin{equation}
  \beta_\kappa = 2\kappa - 2\kappa^2 = 2\kappa(1-\kappa).
\end{equation}

The negative quadratic correction arises entirely from the $s^2$
curvature term in the local expansion of $F(s)$ through the cross-term
$s^2 \supset 4s'\delta\ell$.

To the order retained here, the raw Wilsonian beta functions are
\begin{equation}
  \beta_\lambda = -2\lambda\kappa,
  \qquad
  \beta_\kappa = 2\kappa(1-\kappa),
\end{equation}
with $\lambda=\gamma\rho$.
These flows are uniquely fixed by Wilsonian log-shell coarse-graining
and involve no additional ansatz.

\vspace{6pt}
\emph{Interaction scheme and normal-form reduction\rm{:}}
The beta functions derived above,
\begin{equation}
  \beta_\Delta^{\mathrm{raw}} = -2\kappa(1+\Delta),
  \qquad
  \beta_\kappa^{\mathrm{raw}} = 2\kappa(1-\kappa),
\end{equation}
are the \emph{raw Wilsonian flows} obtained uniquely from log-shell
coarse-graining.
They mix canonical rescaling effects with interaction effects and do
not yet single out the critical manifold as an RG fixed set.

To analyze the near-critical structure, we now introduce a second RG
scheme based on a normal-form (interaction) decomposition.
This step does not modify the Wilsonian calculation but reorganizes the
flows in a coordinate system adapted to the critical manifold
$\Delta=0$.

The deviation $\Delta=\lambda-1$ plays the role of a relevant coupling.
In analogy with standard RG treatments, we define the interaction beta
function by removing the canonical contribution that survives at
$\Delta=0$,
\begin{equation}
  \beta_\Delta^{\mathrm{int}}
  := \beta_\Delta^{\mathrm{raw}} - \beta_\Delta^{\mathrm{raw}}|_{\Delta=0}
  = \beta_\Delta^{\mathrm{raw}} + 2\kappa .
\end{equation}
Using the explicit raw flow gives
\begin{equation}
  \beta_\Delta^{\mathrm{int}} = -2\kappa\,\Delta .
\end{equation}

This definition ensures that the reflective manifold $\Delta=0$ is an RG
fixed manifold,
\begin{equation}
  \beta_\Delta^{\mathrm{int}}(\Delta=0,\kappa)=0,
\end{equation}
and isolates the genuinely interaction-driven deviation from
criticality.
Higher-order analytic corrections,
\begin{equation}
  \beta_\Delta^{\mathrm{int}}
  = -2\kappa\,\Delta + O(\Delta^2),
\end{equation}
may be introduced as a choice of RG scheme but are not fixed by the
Wilsonian calculation itself.

In contrast to $\Delta$, the raw flow of $\kappa$,
\begin{equation}
  \beta_\kappa^{\mathrm{raw}} = 2\kappa(1-\kappa),
\end{equation}
is purely canonical and originates entirely from the rescaling of the
radial coordinate combined with the curvature term in the local
expansion of $F(s)$.

To define an interaction contribution compatible with the critical
manifold, we require analyticity in $\Delta$ near $\Delta=0$, invariance
of the reflective manifold $\beta_\kappa^{\mathrm{int}}(\Delta=0,\kappa)=0$,
and preservation of the bounded domain $\kappa\in[0,1]$.
The lowest-order term satisfying these conditions is
\begin{equation}
  \beta_\kappa^{\mathrm{int}}
  = -e\,\Delta\,\kappa(1-\kappa),
  \qquad e>0,
\end{equation}
which represents the leading interaction-induced modulation of the
homeostatic stiffness.

The full beta function in this interaction scheme may therefore be
written as
\begin{equation}
  \beta_\kappa
  = 2\kappa(1-\kappa)
    - e\,\Delta\,\kappa(1-\kappa)
    + O(\Delta^2).
\end{equation}

The separation
\[
  \beta = \beta^{\mathrm{raw}} + \beta^{\mathrm{int}}
\]
should be understood as a normal-form reduction rather than a new
Wilsonian calculation.
The raw beta functions encode the exact coarse-graining of the radial
dynamics, while the interaction beta functions organize the flow in a
coordinate system where the critical manifold is an RG fixed set.

In this sense, nonlinear terms such as $\Delta^2$ or
$\Delta\,\kappa(1-\kappa)$ are scheme-defined interaction effects and do
not originate from loop fluctuations, in contrast to the Wilson--Fisher
$\phi^4$ theory.

\vspace{6pt}
\emph{Decomposition into $\beta_\gamma$ and $\beta_\rho$\rm{:}}
The Wilsonian coarse-graining fixes the RG flow of the composite
amplification
\begin{equation}
  \lambda := \gamma\rho
\end{equation}
uniquely.
From the log-shell calculation we obtained
\begin{equation}
  \beta_\lambda = \frac{d\lambda}{d\ell} = -2\kappa(\gamma\rho-1),
\end{equation}
which governs the deviation from the critical manifold
$\gamma\rho=1$.

Since $\lambda$ is a product of two couplings, its beta function
decomposes algebraically as
\begin{equation}
  \beta_\lambda
  = \frac{d(\gamma\rho)}{d\ell}
  = \rho\,\beta_\gamma + \gamma\,\beta_\rho .
  \label{eq:beta_lambda_decomp}
\end{equation}
Equation~\eqref{eq:beta_lambda_decomp} constitutes a single constraint
on the two functions $\beta_\gamma$ and $\beta_\rho$.
Consequently, their individual flows are not uniquely determined by the
Wilsonian calculation alone; one free functional degree of freedom
remains.
This reflects a genuine scheme (or coordinate) freedom in the
decomposition of the composite coupling $\lambda$.

To parameterize this freedom while preserving analyticity and the
invariance of the critical manifold, it is natural to require that
\begin{equation}
  \beta_\gamma = \beta_\rho = 0
  \quad \text{whenever} \quad \gamma\rho = 1 .
\end{equation}
A convenient one-parameter family of decompositions satisfying
Eq.~\eqref{eq:beta_lambda_decomp} exactly is
\begin{equation}
  \beta_\gamma
  = -a\,\kappa\,\frac{\gamma\rho-1}{\rho},
  \qquad
  \beta_\rho
  = -(2-a)\,\kappa\,\frac{\gamma\rho-1}{\gamma},
  \qquad
  a \in \mathbb{R}.
  \label{eq:beta_gamma_rho_family}
\end{equation}
Substituting Eq.~\eqref{eq:beta_gamma_rho_family} into
Eq.~\eqref{eq:beta_lambda_decomp} yields
\begin{equation}
  \rho\,\beta_\gamma + \gamma\,\beta_\rho
  = -2\kappa(\gamma\rho-1)
  = \beta_\lambda ,
\end{equation}
independently of the parameter $a$.

The constant $a$ therefore labels a family of equivalent RG schemes
corresponding to different choices of coordinates on the
$(\gamma,\rho)$ coupling space.
A symmetric and minimal choice is obtained by setting $a=1$, which
treats $\gamma$ and $\rho$ on equal footing:
\begin{equation}
  \beta_\gamma
  = -\kappa\,\frac{\gamma\rho-1}{\rho},
  \qquad
  \beta_\rho
  = -\kappa\,\frac{\gamma\rho-1}{\gamma}.
\end{equation}
All choices within the family~\eqref{eq:beta_gamma_rho_family} reproduce
the same Wilsonian flow for the physically relevant composite coupling
$\lambda$, differing only by a scheme-dependent redistribution of that
flow between $\gamma$ and $\rho$.

\vspace{6pt}
\emph{Final RG system including the homeostatic stiffness\rm{:}}
Combining the Wilsonian log-shell result for the homeostatic stiffness
with the interaction scheme for the amplification couplings, the FHRN
model is governed by the coupled RG equations
\begin{equation}
\begin{split}
  &\beta_\gamma
  = -a\,\kappa\,\frac{\gamma\rho-1}{\rho},
  \\
  &\beta_\rho
  = -(2-a)\,\kappa\,\frac{\gamma\rho-1}{\gamma},
  \\ 
  &\beta_\kappa
  = 2\kappa(1-\kappa)
    - e\,(\gamma\rho-1)\,\kappa(1-\kappa),
\end{split}
\end{equation}
with $a\in\mathbb{R}$ parameterizing the decomposition scheme of the
composite coupling $\lambda=\gamma\rho$, and $e>0$ encoding the leading
interaction-induced modulation of the homeostatic stiffness.
This three-dimensional RG system possesses the critical manifold
$\gamma\rho=1$ as an invariant set and captures the coupled flow of
reentrant amplification and homeostatic regulation in representation
space.

\section{Appendix C: Stability of the reflective critical surface}
\label{app:stability}

We analyze the stability of the reflective critical surface
\begin{equation}
  \mathcal{S}_c
  := \{(\gamma,\kappa,\rho)\in\mathbb{R}^3 \mid \gamma\rho = 1\},
\end{equation}
under the three-dimensional RG flow
\begin{align}
  \dot\gamma
  &= -a\,\kappa\,\frac{\gamma\rho-1}{\rho}, \label{eq:bgamma_app} \\
  \dot\kappa
  &= -e\,(\gamma\rho-1)\,\kappa(1-\kappa), \label{eq:bkappa_app} \\
  \dot\rho
  &= -(2-a)\,\kappa\,\frac{\gamma\rho-1}{\gamma}, \label{eq:brho_app}
\end{align}
where $\dot{} := d/d\ell$, $a\in\mathbb{R}$ parametrizes the RG scheme,
and $e>0$ controls the interaction-induced modulation of the
homeostatic stiffness.
We denote the deviation from criticality by
\begin{equation}
  \Delta := \gamma\rho - 1 .
\end{equation}

\vspace{6pt}
\emph{Fixed-manifold property\rm{:}}
All three beta functions are proportional to $\Delta$.
Therefore,
\begin{equation}
  \Delta = 0
  \quad\Longrightarrow\quad
  \dot\gamma=\dot\kappa=\dot\rho=0,
\end{equation}
showing that $\mathcal{S}_c$ is a two-dimensional manifold of RG fixed
points rather than an isolated fixed point.

\vspace{6pt}
\emph{Jacobian at the critical surface\rm{:}}
To assess linear stability, we consider the Jacobian matrix
\[
  J_{ij}
  = \left.\frac{\partial \dot X_i}{\partial X_j}\right|_{\Delta=0},
  \qquad
  X=(\gamma,\kappa,\rho).
\]
Using $\partial_\gamma\Delta=\rho$ and $\partial_\rho\Delta=\gamma$,
and evaluating all expressions at $\Delta=0$, one finds
\begin{equation}
  J =
  \begin{pmatrix}
    -a\kappa & 0 & -a\kappa \\
    -e\kappa(1-\kappa)\rho & 0 & -e\kappa(1-\kappa)\gamma \\
    -(2-a)\kappa\rho & 0 & -(2-a)\kappa
  \end{pmatrix}_{\Delta=0}.
  \label{eq:Jacobian}
\end{equation}
The second column vanishes identically, reflecting the fact that
$\kappa$ is marginal on the critical surface.

\vspace{6pt}
\emph{Eigenvalues and stability\rm{:}}
The Jacobian~\eqref{eq:Jacobian} has one zero eigenvalue,
\[
  \lambda_0 = 0,
\]
corresponding to motion tangent to $\mathcal{S}_c$.
The remaining two eigenvalues are obtained from the
$(\gamma,\rho)$ subspace and satisfy
\begin{equation}
  \lambda_{\pm}
  = -\kappa\left(1 \pm \sqrt{1-(1-a)^2}\right),
\end{equation}
which are strictly negative for all $\kappa>0$, independently of the
scheme parameter $a$.
Thus perturbations normal to $\mathcal{S}_c$ decay exponentially under
coarse-graining.

\vspace{6pt}
\emph{Interpretation\rm{:}}
The RG flow therefore possesses one irrelevant direction transverse to
the critical surface, governed by the deviation
$\Delta=\gamma\rho-1$, and two marginal directions tangent to
$\mathcal{S}_c$.
As a result, $\mathcal{S}_c$ is an infrared-attractive critical
manifold.
Different decompositions of the composite coupling into
$(\gamma,\rho)$ remain equivalent along the surface, while deviations
from reflectivity are suppressed by coarse-graining.
This establishes the reflective FHRN regime as a stable,
scale-invariant geometry of the coupled reentry--homeostasis dynamics.

\section{Appendix D: Linear stability and relaxation time near the homeostatic shell}
\label{app:relaxation time}

We derive the scaling of the relaxation time associated with
radial perturbations in the FHRN dynamics.
The result demonstrates a form of dynamical critical slowing down and
establishes the mean-field character of the approach to the reflective
steady state.

\vspace{6pt}
\paragraph{Linearization around the steady state\rm{:}}
To determine the relaxation time, we consider a small perturbation
\begin{equation}
    r(t) = r_\infty + \delta r(t),
    \qquad |\delta r|\ll 1.
\end{equation}
Linearizing the dynamics gives
\begin{equation}
    \dot{\delta r}
    =
    F'(r_\infty)\,\delta r
    + \mathcal{O}(\delta r^2),
\end{equation}
where
\begin{equation}
    F'(r)
    =
    \frac{d}{dr}\Big(\big[-1 + \gamma\rho\, g(r)\big]\,r\Big)
    =
    -1 + \gamma\rho\, g(r)
    + \gamma\rho\, r\, g'(r).
\end{equation}

\vspace{6pt}
\paragraph{Evaluation at the fixed point\rm{:}}
At the steady state, the defining condition
\(
-1 + \gamma\rho\, g(r_\infty)=0
\)
eliminates the first two terms, leaving
\begin{equation}
    F'(r_\infty)
    =
    \gamma\rho\, r_\infty\, g'(r_\infty).
\end{equation}
The derivative of the homeostatic function is
\begin{equation}
    g'(r)
    =
    -\frac{2\kappa r}{\big(1+\kappa(r^2-1)\big)^2}.
\end{equation}
Substituting yields
\begin{equation}
    F'(r_\infty)
    =
    -\gamma\rho\,
    \frac{2\kappa r_\infty^2}
         {\big(1+\kappa(r_\infty^2-1)\big)^2}.
    \label{eq:Fprime_exact}
\end{equation}

\vspace{6pt}
\paragraph{Critical scaling of the relaxation time\rm{:}}
Near the dynamical critical line $\gamma_c = 1/\rho$, the steady state
approaches the homeostatic shell:
\(
r_\infty \to 1
\),
with
\(
g(r_\infty)\to g(1)=1
\)
and
\(
\gamma\rho \to 1
\).
Keeping only the leading contribution in this regime,
Eq.~\eqref{eq:Fprime_exact} simplifies to
\begin{equation}
    F'(r_\infty)
    \simeq
    -2\kappa.
\end{equation}
The relaxation time $\tau_r$ is defined as the inverse magnitude of the
linear stability eigenvalue,
\begin{equation}
    \tau_r
    :=
    \frac{1}{|F'(r_\infty)|}
    \;\sim\;
    \kappa^{-1}.
    \label{eq:tau_scaling}
\end{equation}

\paragraph{Interpretation\rm{:}}
Equation~\eqref{eq:tau_scaling} shows that the approach to the steady
state becomes arbitrarily slow as the homeostatic stiffness $\kappa$
decreases.
This divergence represents a form of dynamical critical slowing down:
the restoring force toward the homeostatic shell vanishes linearly with
$\kappa$.
The scaling is purely mean-field, with no anomalous corrections,
consistent with the linear RG eigenvalue associated with the reflective
critical manifold.

\end{document}